\documentclass[namedreferences]{kluwer}
\usepackage[dvips]{graphicx}
\begin{document}
\begin{article}
\begin{opening}
\title{Low-Luminosity Accretion in Black Hole X-ray Binaries and
Active Galactic Nuclei}

%\subtitle{}

\author{Ramesh \surname{Narayan}\email{narayan@cfa.harvard.edu}}
\institute{Harvard-Smithsonian Center for Astrophysics, 60 Garden
Street, Cambridge, MA 02138, U.S.A.}

%\date: rather not

%\dedication{To Jim}

%\translation{De Kluwer LaTeX stylefile; aanwijzingen voor auteurs}

\runningtitle{Low-Luminosity Accretion}
\runningauthor{Narayan}

\begin{ao}
Kluwer Prepress Department\\
P.O. Box 990\\
3300 AZ Dordrecht\\
The Netherlands
\end{ao} 

%\begin{motto}
%What can't be done with TeX isn't worth doing.
%\end{motto}

\begin{abstract} 
At luminosities below a few percent of Eddington, accreting black
holes switch to a hard spectral state which is very different from the
soft blackbody-like spectral state that is found at higher
luminosities.  The hard state is well-described by a two-temperature,
optically thin, geometrically thick, advection-dominated accretion
flow (ADAF) in which the ions are extremely hot (up to $10^{12}$ K
near the black hole), the electrons are also hot ($\sim10^{9-10.5}$
K), and thermal Comptonization dominates the X-ray emission.  The
radiative efficiency of an ADAF decreases rapidly with decreasing mass
accretion rate, becoming extremely low when a source reaches
quiescence.  ADAFs are expected to have strong outflows, which may
explain why relativistic jets are often inferred from the radio
emission of these sources.  It has been suggested that most of the
X-ray emission also comes from a jet, but this is less well
established.
\end{abstract}


%\abbreviations{\abbrev{KAP}{Kluwer Academic Publishers};
%   \abbrev{compuscript}{Electronically submitted article}}

%\nomenclature{\nomen{KAP}{Kluwer Academic Publishers};
%   \nomen{compuscript}{Electronically submitted article}}

%\classification{JEL codes}{D24, L60, 047}
\end{opening}
\section{Introduction}
The well-known thin accretion disk model has been a staple of
accretion theory for more than 30 years (Shakura \& Sunyaev 1973;
Novikov \& Thorne 1973).  It provides a good description of the soft
X-ray spectra of luminous black hole X-ray binaries (XRBs) in the high
soft state (see McClintock \& Remillard 2004) and the big blue bump in
the optical/UV spectra of bright quasars and active galactic nuclei
(AGN; Malkan 1983; but see Koratkar \& Blaes 1999).  However, even
from the earliest days (Tananbaum et al. 1972) it was realized that
XRBs sometimes switch to a hard spectral state which requires the
accreting gas to be hot and optically thin, quite different from the
gas in a thin disk which is relatively cool and optically thick.

Observations of a number of XRBs have shown that, at luminosities
below a few percent of Eddington, the sources enter the classic low
hard state, and at much lower luminosities the quiescent state
(McClintock \& Remillard 2004).  Both states are characterized by very
high temperatures $\sim100$ keV or more, optically thin emission, and
weak or absent soft X-ray emission.  In the case of supermassive black
holes, low-luminosity AGN (LLAGN) are noted for the absence of a big
blue bump and the presence of substantial hard X-ray and radio
emission (Ho 1999; Quataert et al. 1999; Nagar et al. 2000).  This
again indicates that a standard thin accretion disk is either absent
or is energetically unimportant, and that a hot flow, similar to those
seen in low-luminosity XRBs, is probably present.

In an important paper, Shapiro, Lightman \& Eardley (1976) introduced
the idea of a two-temperature plasma and used it to develop a new hot
accretion solution which is distinct from the standard thin disk.
However, the solution turned out to be thermally unstable (Pringle
1976).  Fortunately, there is a second hot two-temperature solution
called an advection-dominated accretion flow (ADAF; Narayan \& Yi
1994, 1995b; Abramowicz et al. 1995; see Narayan, Mahadevan \&
Quataert 1978; Kato, Fukue \& Mineshige 1998 for reviews).  This
solution, which is also referred to as a radiatively inefficient
accretion flow (RIAF), was originally discussed in a forgotten paper
by Ichimaru (1977; see also Rees et al. 1982).  It has been shown to
be effectively stable (Kato et al. 1997; Wu 1997), and it is now
recognized to be relevant for understanding low-luminosity accretion
flows around black holes.

\section{Advection-Dominated Accretion Flow}

\subsection{ Basic Properties}

The energy equation of gas in a time-steady accretion disk may be
written schematically as
\begin{equation}
q_{\rm adv} \equiv \rho v {Tds\over dR} = q_+ - q_-,
\end{equation}
where $q_{\rm adv}$ is the rate of advection of energy per unit
volume, $\rho$ is the density, $v$ is the radial velocity, $T$ is the
temperature, $s$ is the specific entropy, $R$ is the radius, $q_+$ is
the viscous heating rate per unit volume, and $q_-$ is the radiative
cooling rate.  A thin accretion disk is characterized by the condition
$q_+ \sim q_- \gg q_{\rm adv}$, i.e., viscous heating is balanced by
radiative cooling.  In contrast, an ADAF satisfies $q_+ \sim q_{\rm
adv} \gg q_-$, i.e., most of the viscous heat remains trapped in the
gas (because the gas is radiatively inefficient), and the energy is
advected in towards the BH.  Technically, since the plasma is
two-temperature, it is necessary to write separate energy equations
for the ions and the electrons and to model the energy transfer
between the two species by Coulomb collisions (Narayan \& Yi 1995b;
Nakamura et al. 1997).  We do not go into the details here and refer
the reader to the review by Narayan et al. (1998).

The ADAF solution has a number of interesting properties:  

\noindent
(i) The ion temperature varies roughly as $T_i \sim 10^{12}{\rm K}/r$,
where $r=R/R_S$ is the radius in Schwarzschild units.  The electron
temperature, however, saturates at $T_e\sim10^9-10^{10.5}$K for
$r\lesssim10^2-10^3$.  

\noindent
(ii) The large ion temperature implies that the flow is geometrically
thick.  In fact, an ADAF might be viewed as the viscous rotating
analog of spherical Bondi accretion.

\noindent
(iii) The gas in an ADAF is optically thin; therefore, the radiation
from the hot electrons (which dominate the emission) is primarily by
thermal Comptonization. Because Comptonization acts as a natural
thermostat, the electron temperature is typically $\sim 100$ to a few
100 keV and varies by only a factor of a few over a wide range of
Eddington-scaled accretion rate $\dot m = \dot M/\dot M_{\rm Edd}$
(e.g., Esin, McClintock \& Narayan 1997; Esin et al. 1998; Zdziarski
et al. 2003).

\noindent
(iv) The ADAF solution exists only for accretion rates $\dot m$ below
a certain critical rate $\dot m_{\rm crit}$, whose value depends on
the viscosity parameter $\alpha$.  For $\alpha \sim0.1-0.25$, $\dot
m_{\rm crit} \sim0.01-0.1$.

\subsection{Application of the ADAF Solution to XRBs and AGN}

Rather miraculously, the properties of the ADAF solution are exactly
what are needed to understand XRBs at low luminosities.  The radiation
in an ADAF is dominated by thermal Comptonization, in agreement with
observations of XRBs in the low hard state.  The electron temperature
is about 100 keV, exactly what is needed to explain X-ray spectra in
the low state (e.g., see the ADAF model of GRO J0422+32 shown in
Fig. 1).  Finally, the critical $\dot m_{\rm crit}\sim0.01-0.1$ above
which the ADAF solution ceases to exist is consistent with the
luminosity at which the transition from the low hard state to the high
soft state occurs in XRBs.

\begin{figure}
\centerline{\includegraphics[width=20pc]{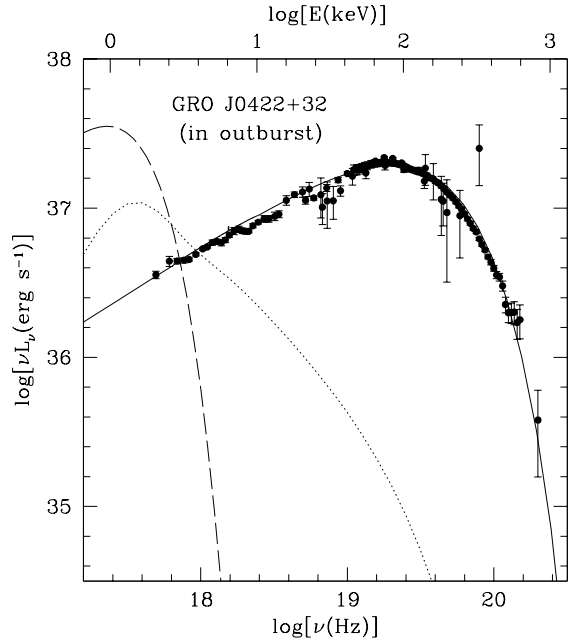}}
\caption{Combined TTM (2--20 keV), HEXE (20--200 keV), and OSSE
(50--600 keV) spectrum of GRO J0422+32 in the low hard state.  The
solid line shows an ADAF fit to the spectrum.  (From Esin et
al. 1998)}
\end{figure}

By combining the thin accretion disk model and the ADAF model, Narayan
(1996) and Esin et al. (1997) showed that it is possible to understand
qualitatively the various spectral states of XRBs.  According to their
proposal (Fig. 2), for $\dot m > \dot m_{\rm crit}$, the accretion
occurs primarily via a thin disk with a corona on top.  This
corresponds to the high soft state, with the disk providing the bulk
of the radiation via a multicolor blackbody component and the corona
contributing hard X-rays through Compton scattering (Haardt \&
Maraschi 1991).  Once $\dot m$ falls below $\dot m_{\rm crit}$, a hole
opens up at the center of the disk and the hole is filled with a hot
ADAF.  For $\dot m \lesssim \dot m_{\rm crit}$, the hole is relatively
small and both the thin disk and the ADAF contribute roughly equally.
This corresponds to the intermediate state.  With decreasing $\dot m$,
the transition radius $r_{\rm tr}$ between the two zones becomes
larger and the ADAF dominates the energetics.  At very low $\dot m$,
i.e., in the quiescent state, $r_{\rm tr}$ is very large ($>1000$,
e.g., Narayan, McClintock \& Yi 1996); in objects such as Sgr A*
(Yuan, Quataert \& Narayan 2003), the outer thin disk may even
disappear altogether.

\begin{figure}
\centerline{\includegraphics[width=20pc]{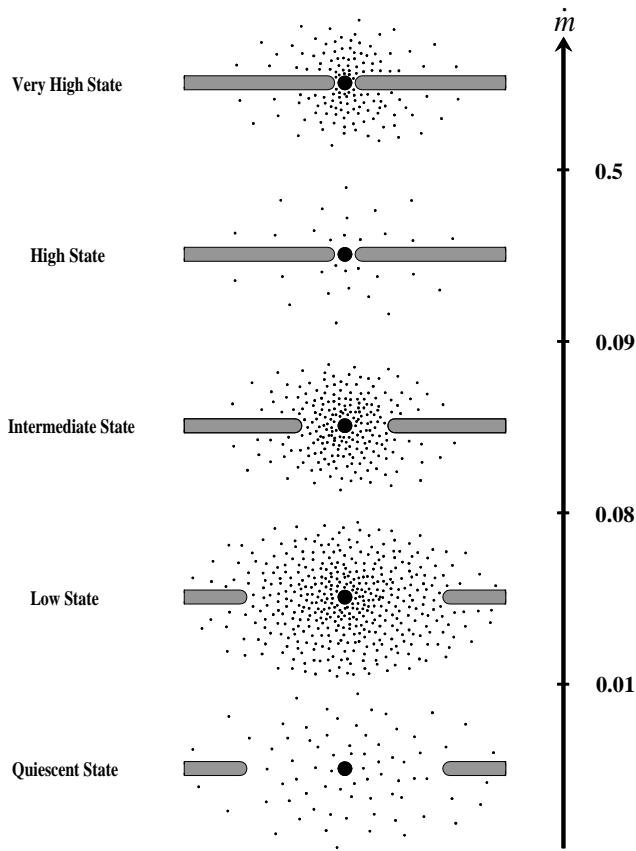}}
\caption{Configuration of the accretion flow around a black hole in
different spectral states, shown schematically as a function of the
Eddington-scaled mass accretion rate $\dot m$.  The ADAF and the
corona are indicated by dots and the thin disk by the shaded
horizontal bars.  (From Esin et al. 1997)}
\end{figure}

In the intermediate state and at the high end of the low state, the
ADAF is only mildly advection-dominated, so the radiative efficiency
is fairly large.  However, with decreasing $\dot m$, the efficiency
drops rapidly.  Quiescent systems are, therefore, radiatively very
inefficient; Sgr A*, for instance, has a luminosity that is only about
$10^{-5}-10^{-6}$ of the rate at which rest mass energy accretes from
its surroundings (Yuan et al. 2003).

An ADAF has two sources of soft photons for Comptonization, and both
are included in models (Narayan, Barret \& McClintock 1997): (i)
thermal synchrotron photons from the hot electrons in the ADAF, (ii)
thermal blackbody photons from the outer thin disk.  The former
dominates at low $\dot m$ (quiescent state and lower end of low state)
and the latter at higher $\dot m$ (upper end of low state and
intermediate state).

The above paradigm, which is based on the ADAF model or its variants
(ADIOS, Blandford \& Begelman 1999; LHAF, Yuan 2001, Yuan \& Zdziarski
2004; CDAF, Narayan, Igumenshchev \& Abramowicz 2000, Quataert \&
Gruzinov 2000), explains qualitatively many observations of XRBs (Esin
et al. 1997, 1998, 2001; see Narayan et al. 1998 for other
applications).  Very recently, Meyer-Hofmeister, Liu \& Meyer (2004)
have suggested an interesting mechanism involving an interplay between
Compton cooling and disk evaporation to explain the hysteresis
phenomenon that has been identified in the high-to-low transition of
black hole and neutron star X-ray binaries (Miyamoto et al. 1995;
Nowak, Wilms \& Dove 2002; Maccarone \& Coppi 2003; Zdziarski et
al. 2004).

The ADAF model also explains a variety of observations of LLAGN: Sgr
A* (Narayan, Yi \& Mahadevan 1995; Yuan et al.  2003), LLAGN in giant
ellipticals (Fabian \& Rees 1995; Reynolds et al. 1996; Di Matteo et
al. 2003), and LINERs (Lasota et al. 1996; Quataert et al. 1999).  In
addition, it appears that ADAFs may be present in BL Lac objects
(Maraschi \& Tavecchio 2003), FR I sources (Reynolds et al.  1996;
Begelman \& Celotti 2004), XBONGs (Yuan \& Narayan 2004), and even
some Seyferts (Chiang \& Blaes 2003).  Overall, the model has turned
out to be quite useful for providing a qualitative understanding of a
variety of phenomena in low-luminosity accreting black holes (Quataert
2001; Narayan 2002).

\subsection{Transition Radius}

A key element of the model shown in Fig. 2 is that the transition
radius $r_{\rm tr}$ between the outer thin disk and the inner ADAF
varies with $\dot m$.  But how exactly does it vary?  To calculate
this  from first principles, one needs a physical theory of what causes
the transition between the two kinds of flow.  A number of ideas have
been discussed in the literature and many efforts have been devoted to
estimating $r_{\rm tr}(\dot m)$ theoretically (Meyer \&
Meyer-Hofmeister 1994; Dullemond \& Turolla 1998; Liu et al. 1999;
Rozanska \& Czerny 2000; Spruit \& Deufel 2002), but no model is
presently able to provide robust predictions.

An alternative approach is to use the observations themselves to
determine $r_{\rm tr}(\dot m)$.  For a number of sources, by fitting
observations one is able to obtain estimates of the Eddington-scaled
luminosity and mass accretion rate, as well as the transition radius.
Although there are large uncertainties in some of these quantities,
nevertheless the results are interesting when plotted, as discussed in
Narayan et al. (1998) and Yuan \& Narayan (2004).  Figure 3 from the
latter paper supports the basic features of the Esin et al. (1997)
proposal; specifically, the transition radius $r_{\rm tr}$ seems to
increase monotonically with decreasing luminosity, as postulated in
the model.  Interestingly, both XRBs and LLAGN are included in the
plot, and the two classes of sources seem to follow more or less the
same trend even though their masses are very different.  This confirms
that the physics of ADAFs is largely mass-independent (Narayan \& Yi
1995b), once all quantities are scaled suitably in terms of Eddington
and Schwarzschild units.

\begin{figure}
\centerline{\includegraphics[width=20pc]{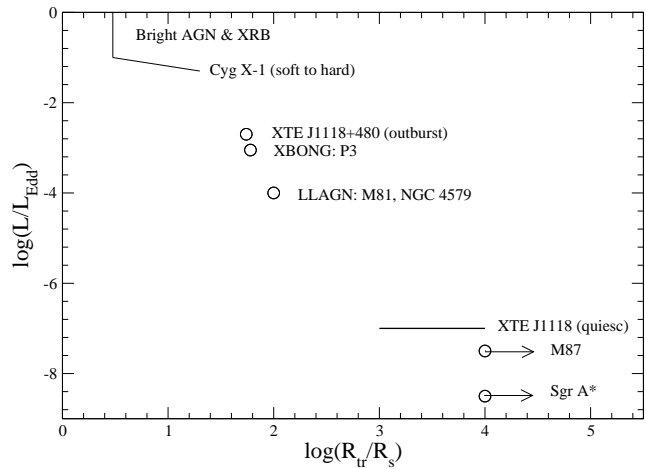}}
\caption{Bolometric luminosity in Eddington units (ordinate) vs. the
transition radius in Schwarzschild units (abscissa) for different
observed systems.  (See Yuan \& Narayan 2004 for details)}
\end{figure}

\section{Role of Jets}

\subsection{ADAFs, Convection, Outflows, Jets}

One of the interesting properties of ADAFs, highlighted already in the
first papers (Narayan \& Yi 1994, 1995a), is that the accreting gas
has a positive Bernoulli parameter, i.e., the gas is technically not
bound to the BH.  One expects, therefore, strong winds and outflows
from an ADAF (Narayan \& Yi 1994, 1995a).  Another property of ADAFs
is that they have unstable entropy gradients and are hence violently
unstable to convection (Begelman \& Meier 1982; Narayan \& Yi 1994,
1995a).  Both effects have been seen in hydrodynamic and MHD
simulations of ADAFs (Stone, Pringle \& Begelman 1999; Igumenshchev \&
Abramowicz 2000; Narayan et al.  2000; Stone \& Pringle 2001; Hawley
\& Balbus 2002; Igumenshchev, Narayan \& Abramowicz 2003).

An important consequence of the above effects is that the mass that
accretes onto the BH via an ADAF is much less than the mass supplied at
the outer edge of the accretion flow.  Blandford \& Begelman (1999)
suggested that the mass accretion rate may scale with radius as $\dot
m \sim (r/r_{\rm out})^s$.  This scaling has been widely used in ADAF
models (e.g., Quataert \& Narayan 1999).  One problem is that the
value of $s$ cannot be estimated from first principles, though one may
be able, in favorable cases, to fit $s$ by comparison to observations.
Yuan et al. (2003) estimated $s \sim 0.3$ for the accretion flow in
Sgr A*.  For this choice of $s$, the unusually low luminosity of the
source is explained partly by the reduced mass accreting on the black
hole ($\sim 10^{-2}$ of the mass available at the Bondi radius) and
partly by the low radiative efficiency of the accreting gas
($\sim10^{-3}$).

Once we recognize that ADAFs have powerful outflows, it is natural to
think that these flows would have relativistic jets (Meier 2001).
Indeed, such a connection has been established fairly convincingly.
XRBs in the low state generally have measurable radio emission,
whereas sources in the high state do not.  The radio emission has been
resolved into a jet in Cyg X--1, and jets are inferred in other
sources because of their large brightness temperatures (Fender 2004).
In the case of supermassive BHs again, it is found that LLAGN are in
general radio loud with high brightness temperatures (Nagar et
al. 2000; Falcke et al. 2000).  Also, BL Lacs, which have been
associated with ADAFs (Maraschi \& Tavecchio 2003), are known to have
strong jets.  Apart from these experimental indications, there is also
a strong theoretical argument for jets, viz., an ADAF simply cannot
produce the large radio fluxes that are observed.  The radio emission
has to come from a volume much larger than the ADAF, which suggests
that it must originate in a jet.

\subsection{Does the Jet Dominate the High Energy Emission in ADAFs?}

While the argument for the radio emission originating in a jet is
clear, what about the X-ray emission?  The ADAF model is quite
successful in explaining the X-ray fluxes and spectra of
low-luminosity black holes without invoking a jet (e.g., Fig. 1).  One
source of particular interest is XTE J1118+480, for which a nearly
complete spectrum has been measured in the low state (McClintock et
al. 2001).  Esin et al. (2001) proposed a model for this source in
which (as in Fig. 2) a thin disk is present outside a transition
radius $r_{\rm tr} \sim 50$ and an ADAF is present inside this radius.
The model fits the spectral data in the optical, UV and X-rays quite
well.

Soon after this work, Markoff et al. (2001) proposed an alternative
model in which they explained the entire spectrum of XTE J1118+480
from radio to X-rays by means of synchrotron emission from a jet.
(They invoked a standard disk for the optical and UV.)  As described
earlier, a jet is certainly expected in an ADAF system and it is quite
natural for the jet to dominate in radio and perhaps infrared.  What
was surprising was that the Markoff et al. model was able to explain
the X-ray emission with the same jet.

The case for a jet became stronger when Corbel et al. (2003) showed
that there is a strong correlation between the radio and X-ray
emission in the black hole XRB GX 339--4 in the low state and
quiescent state.  They suggested that a significant fraction of the
X-ray emission may originate in a jet.  Interestingly, both Markoff et
al. and Corbel et al. require a radiatively inefficient ADAF to be
present since a radiatively efficient disk would swamp the jet
emission in their model.  However, the ADAF is postulated to be
virtually silent even in the X-ray band, and it is the jet that
produces most of the observed radiation.

Heinz \& Sunyaev (2003) studied the jet model and worked out a scaling
relation between the synchrotron flux at a given frequency, the mass
of the black hole, and the mass accretion rate.  Their model is
applicable to jets anchored in either an ADAF or a standard disk.
Merloni, Heinz \& di Matteo (2003) extended this work and showed that
accreting black holes follow quite well a ``fundamental plane'' in the
three-dimensional parameter space of radio luminosity, X-ray
luminosity and black hole mass.  However, they came down in favor of
the ADAF rather than the jet as the source of the X-ray emission in
the low hard state.  Falcke, K\"ording \& Markoff (2004) argued
instead that synchrotron emission from the jet is the source of the
X-rays.  In a recent paper, Heinz (2004) has presented additional
arguments why a synchrotron jet is unlikely to explain the X-ray
emission in low hard state binaries.

Apart from the above contradictory arguments, Zdziarski et al. (2003)
have presented possible additional difficulties with a jet
interpretation of the X-ray emission in the low state.  They claim
that synchrotron emission cannot produce as sharp a cut off at high
energies as observed (e.g., see the spectrum shown in Fig. 1).  Also,
the predicted spectrum is not as hard as the spectra observed in some
low state XRBs.  Finally, the fact that the cutoff occurs near 100 or
a few 100 keV in several sources (in fact, all sources in which a
cutoff has been seen) does not find a natural explanation in the jet
model; it requires a degree of fine-tuning of the power-law energy
distribution of the radiating electrons.  In the ADAF model, on the
other hand, thermal Comptonization acts as a thermostat that naturally
produces a temperature on the order of 100 keV.

Recently, Yuan et al. (2004) and Malzac et al. (2004) have come up
with a jet-ADAF model of J1118+480 in which a jet produces most of the
radio and infrared emission, the ADAF produces the X-ray emission, and
the outer thin disk produces the optical and UV emission.  The model
fits the spectral data satisfactorily and also explains the timing and
variability data qualitatively.  By combining the best features of the
ADAF model and the jet model, this work appears to represent an
interesting compromise between the two models.

\subsection{Is the X-ray Emission Beamed?}

The most obvious feature of a jet is that it involves outward motion
of gas at relativistic speeds.  It is therefore natural to expect
evidence in the data for relativistic beaming.  Observations, however,
generally indicate that beaming is not very pronounced.  Gallo, Fender
\& Pooley (2003) and Fender et al. (2004) estimated an upper limit of
$\gamma\sim2$ for the Lorentz factor of the radio-emitting material,
while Maccarone (2003) deduced $\gamma \leq 1.4$ for the
X-ray-emitting gas in GRO J1655--40.

\begin{figure}
\centerline{\includegraphics[width=20pc]{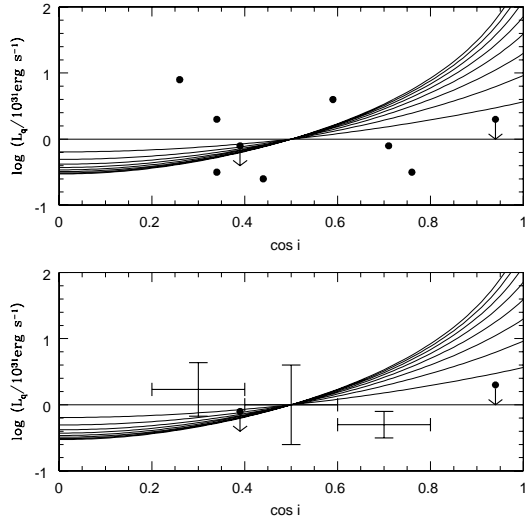}}
\caption{Upper panel: X-ray luminosities of quiescent black hole XRBs
in units of $10^{31} ~{\rm erg\,s^{-1}}$ plotted against the cosine of
the inclination angle $i$.  The curves indicate the expected variation
according to a jet model for different choices of the jet Lorentz
factor: $\gamma = 1.0$ (horizontal line), 1.2, 1.4, ... , 2.6.  Lower
panel: The same data grouped into bins of width 0.2 in $\cos i$.
(From Narayan \& McClintock 2005)}
\end{figure}

Figure 4 shows analogous results for the quiescent state of XRBs.
Assuming the X-ray emission is from a jet, the different curves show
the expected variation of the observed flux as a function of the
binary inclination $i$ for different choices of the jet Lorentz factor
$\gamma$.  The calculations assume that the jet is oriented
perpendicular to the binary orbit, and the curves have been normalized
so as to have the same flux density for $\cos i=0.5$.  Overplotted on
the curves are the quiescent X-ray luminosities of a number of black
hole XRBs.  We see that there is no hint in the data for any increase
in the observed luminosity of low-inclination systems ($\cos i \to
1$).  In fact, the most pole-on system in the sample, 4U1543--47, with
$i\approx 21^o$, has a 95\% confidence upper limit on its quiescent
luminosity that is below the predicted luminosity for all reasonable
values of $\gamma$.  By visual inspection we conclude that, if the
X-ray emission is from a jet, then the Lorentz factor is limited to
$\gamma \lesssim 1.2$.  A likely explanation for the data is that the
X-ray emission is not primarily from an outflowing jet but from an
orbiting ADAF.  Note, however, that the argument assumes the jet to be
oriented perpendicular to the accretion disk which is not supported by
the limited data available (see Narayan \& McClintock 2005).

Jet models generally have radio emission coming from farther out in
the jet and X-ray emission from closer to the center.  In fact, often
the X-rays are postulated to be emitted from the ``base of the jet.''
In this context, we should note that the base of the jet is probably
right inside the ADAF.  It then becomes a matter of semantics whether
this gas should be called the jet or the ADAF.  If the gas were moving
rapidly away from the accreting gas and we could see clear evidence
for beaming, then we could of course confidently claim that it is a
jet.  However, as mentioned above, there is no evidence yet for
relativistic beaming in either the low state or the quiescent state.

The other distinguishing feature of the jet model is that it invokes
synchrotron radiation from nonthermal electrons to explain the X-ray
emission.  The ADAF model, in contrast, makes use of thermal
Comptonization.  As mentioned above, the synchrotron model has some
difficulty explaining certain aspects of the X-ray spectrum.
Nevertheless, the arguments are probably not insurmountable, so the
model must be considered viable.  If, however, it turns out that the
synchrotron idea cannot be made to work for the X-ray emission, and if
one needs to invoke something like thermal Comptonization in the jet
to explain the data, then the argument for the jet would be
significantly weakened.  Why refer to it as a jet if the gas is
located inside the ADAF, is not moving rapidly, and has all the
characteristics of the hot gas in an ADAF?  For all practical
purposes, such a model would be identical to the jet-ADAF model of
Yuan et al. (2004) and Malzac et al. (2004) in which the low-energy
radio (and infrared) emission comes from a bona fide jet, but the high
energy X-ray emission comes from an ADAF.  (Malzac et al. also discuss
the possibility that the X-rays may come from a patchy corona rather
than a standard ADAF.)

One issue still remains to be addressed, viz., the Corbel et
al. (2003) correlation between the radio and X-ray emission.  This
correlation finds a natural explanation in the jet model, but is not
so obvious if the X-rays originate in an ADAF.  If the jet is part of
the general outflow from the ADAF as we have suggested above, then it
is conceivable that there would be a correlation between the
properties of the jet and those of the ADAF.  In this case, even
though the radio and the X-rays come from different parts of the
system (jet and ADAF, respectively), there might still be a strong
correlation between the two.  This possibility is discussed in Meier
(2001) and needs to be investigated quantitatively.

\medskip
The author thanks Tom Maccarone for comments.  This work was supported
in part by grants NAG5-10780 from NASA and AST 0307433 from NSF.

\end{article}
\end{document}